# A negative group delay model for feedback-delayed manual tracking

Revised 6/22/2016  14:18:00

Henning U. Voss
*Weill Cornell Medical College, Citigroup Biomedical Imaging Center*
*516 East 72nd Street, New York, NY10021, USA*
*e-mail address: hev2006@med.cornell.edu*

Nigel Stepp
*HRL Laboratories, LLC, Information Science and Systems Lab*

We propose that feedback-delayed manual tracking performance is limited by fundamental constraints imposed by the physics of negative group delay. To test this hypothesis, the results of an experiment in which subjects demonstrate both reactive and predictive dynamics are modeled by a linear system with delay-induced negative group delay. Although one of the simplest real-time predictors conceivable, this model explains key components of experimental observations. Most notably, it explains the observation that prediction time linearly increases with feedback delay, up to a certain point when tracking performance deteriorates. It also explains the transition from reactive to predictive behavior with increasing feedback delay. The model contains only one free parameter, the feedback gain, which has been fixed by comparison with one set of experimental observations for the reactive case. Our model provides quantitative predictions that can be tested in further experiments.

**1 Introduction**

In feedback-delayed manual tracking experiments tracking performance depends on the subject's ability to predict or anticipate the target positions. This prediction is necessary to overcome the feedback delay time. For example, if the subject is required to track a moving target on a screen, using a stylus on a pad, and there is an artificial delay between the stylus coordinates and the corresponding tracking marker coordinates on the screen, the subject needs to predict in real-time the target position with the stylus position. Therefore, in order to successfully track the target, the stylus position has to lead, or anticipate, the target position on the screen. Besides being of importance for man-machine and computer-machine interaction [1-3], the dynamic and cognitive mechanisms behind delayed manual tracking performance have been an ongoing topic of research [4-8] with possible clinical applications [9,10].

In a recent experiment with inherently chaotic targets [11], one of us has shown that subjects can track the target with feedback delays of up to 400 ms well. The following specific observations about tracking performance have been made:

(i) The actually measured anticipation time, as determined by the argument of the maximum of the cross-correlation function (CCF) between stylus and target, usually is significantly smaller than the feedback delay time.

(ii) If subjects track with a certain anticipation time for a feedback delay of 400 ms, then in other experimental runs with feedback delays of only 200 ms, subjects do not utilize the full prediction performance demonstrated in the previous run but usually use a shorter anticipation time. In other words, participants predict by an amount relative to the given feedback delay, not by the greatest of their ability.

(iii) The measured anticipation time roughly increases linearly with the given feedback delay for up to a critical delay where tracking performance deteriorates.

(iv) Tracking performance as measured by the magnitude of the CCF between stylus and target decreases with feedback delay up to that critical delay.

(v) For small feedback delay, the target is not anticipated but trailed by the stylus. In other words, the subject's behavior changes from predictive to reactive.

We present a mechanism that is based on the physics of systems with negative group delay (NGD) and can explain all of these observations. NGD is known from simple electronic circuits [12,13], but to the best of our knowledge has not been used in the psychological and neurosciences so far. Specifically, the here presented particular form of NGD, *delay-induced* NGD, and its potential relevance in human motor control is not widely known.

First, the model is described. Then the model is simulated and compared with experimental data. Finally, a discussion concludes this paper.

**2 Model description**

Our model is a linear relaxation system with external input and a linear time-delayed feedback [14], i.e.,

$$\dot{y}(t) = -a\,y(t) + b\,x(t) - c\,y(t-\tau)\,, \qquad (1)$$

where $a \geq 0$ is the relaxation coefficient, $x(t)$ the input signal (zero-mean, stationary), $b > 0$ the input scaling, $c \geq 0$ the feedback gain, and $\tau > 0$ the feedback delay.

The meaning of the variables in the delayed manual tracking experiment [11] is as follows: The variable $x$ is the horizontal coordinate of the target on the screen, after a linear normalizing transformation. The variable $y$ is the normalized horizontal stylus coordinate. The variable $y_\tau$ is the delayed and normalized horizontal stylus coordinate of the marker on the screen, which is supposed to continuously track the target marker at position $x$. Therefore, in order to follow the target position $x$, subjects have to continuously predict where its position will be at a time $\tau$ later. In other words, delayed tracking amounts to fulfilling the equation $x = y_\tau$.

Despite its simplicity, it is shown in the following that model (1) has predictive, or anticipatory, as well as reactive properties, which are given by the sign of its group delay. In addition, prediction horizons and stability limits of the model will be derived.

Equation (1) is linear and thus can be described by its frequency response function

$$H(\omega) = \frac{b}{a + i\omega + ce^{-i\omega\tau}} = \frac{b}{\beta}(\beta_1 + i\beta_2), \quad (2)$$

with $\beta_1 = a + c\cos(\omega\tau)$, $\beta_2 = c\sin(\omega\tau) - \omega$, $\beta = \beta_1^2 + \beta_2^2$ [14]. It defines the input/output relationship between $x$ and $y$ under steady-state conditions (after transients have died out) in Fourier space as $Y(\omega) = H(\omega)X(\omega)$, where $f$ is frequency, $\omega = 2\pi f$, $x(t) = \int X(\omega)e^{i\omega t}d\omega$, and $y(t) = \int Y(\omega)e^{i\omega t}d\omega$. If written as $H(\omega) = G(\omega)e^{i\Phi(\omega)}$, its gain is $G(\omega) = b/\sqrt{\beta}$, and its phase is $\Phi(\omega) = \arg(\beta_1 + i\beta_2)$. The group delay is frequency dependent and given by

$$\delta(\omega) = -\frac{d\Phi(\omega)}{d\omega}$$
$$= \frac{c\cos(\omega\tau) - c^2\tau - a(c\tau\cos(\omega\tau) - 1) + c\tau\omega\sin(\omega\tau)}{\beta}. \quad (3)$$

Negative group delay in general means a group advance [12,13], or real-time prediction of the input signal $x$. For the prediction of smooth, i.e., band limited signals, of particular interest is the value of $\delta(\omega)$ for small $\omega$: Expansion in terms of small $\omega$ reveals that there are no linear terms in $\omega$, and if quadratic and higher order terms are neglected in both the counter and denominator of Eq. (3), it follows

$$\delta_0 = \frac{1 - c\tau}{a + c}. \quad (4)$$

This expression has four important consequences for input signal components with small $\omega$:

1. Equation (4) allows both for positive and negative group delay. The group delay is positive for $c\tau < 1$, corresponding to a trailing response. For $c\tau > 1$, the *group delay $\delta_0$ is negative,* a necessary condition for anticipation.

2. The group delay $\delta_0$ is *independent of $\omega$,* a necessary condition for distortion-free signal transfer [13,15].

3. For arbitrary feedback delay $\tau$, the maximum NGD generally is achieved for $a = c$. This follows from Eq. (4) and the fact that the delay-independent (or Lyapunov-) stability of the differential delay equation (1) with $b = 0$ requires $a > c$ [16-18]. Therefore, the maximum prediction horizon is achieved exactly at the margin of instability.

4. The maximum attainable NGD, or prediction horizon, given arbitrary feedback delay, is achieved for $a = c$ in the limit of infinite $a$. It is

$$|\min(\delta_0)| = \tau/2. \quad (5)$$

This value provides a universal upper bound for the prediction horizon for delay-induced NGD processes (1) that cannot be overcome by any choice of parameters, a quite telling constraint when taken in terms of empirical results, below.

The non-dissipative system with $a = 0$ still can have NGD. However, its stability is delay-dependent because the criterion $a > c$ cannot be fulfilled anymore, and a more detailed analysis reveals that the limit obtained in (5) cannot be overcome. It is noted again that these model predictions hold only for input signal components with small $\omega$. Also, only steady-state solutions without transients are captured by the frequency response function. Although of potential interest for a more detailed comparison with experimental data, modeling of transients is not necessary for our purposes and would require an analysis that is beyond the scope of this paper.

Before applying the model, we simplify it further until all parametric freedom is eliminated: First, the model is used with $a = c$, at the margin of stability, in order to minimize the group delay $\delta_0$, Eq. (4). Second, the input scaling $b$ just affects the overall gain and transients and otherwise is of not much importance for the interpretation of the model. We use $b = 2c$, which scales the response $y$ correctly for comparison with $x$ for most stable dynamics observed. Third, the feedback gain $c$ is determined by a fit to experimental observations: Fig. 1c in [11] shows that the errors on the lag/lead times across subjects and sessions are lowest for the smallest experimentally tested feedback delay of $\tau = 20$ ms. Therefore, this delay is used for a fit of $c$ to observations. This case causes trailing behavior on average with an average $\delta_0 = +29$ ms. Inserting this value into Eq. (4) gives $c_0 = 12.82$ s$^{-1}$. Finally, the simplified model is

$$\tau_0 \dot{y}(t) = 2x(t) - y(t) - y(t - \tau), \quad (6)$$

with $\tau_0 = 1/c_0 = 78$ ms. We call the parameter $\tau_0$ the "zero lag time" for reasons that will become apparent just below. Re-insertion of $c_0$ into Eq. (4) for variable delays provides the expected group delays listed in Table 1, second column. As one can see, fixing parameter $\tau_0$ on the case with $\tau = 20$ ms, i.e., trailing behavior, still allows for anticipatory behavior in our model for the larger feedback delays, 200 to 1000 ms.

| Feedback delay $\tau$ | Expected group delay $\delta_0$ | Simulated group delay | Experimental group delay |
|---|---|---|---|
| 20 ms | + 29 ms | + 28 ms | + 29 ms |
| 200 ms | - 61 ms | - 58 ms | - 84 ms |
| 400 ms | - 161 ms | - 156 ms | - 140 ms |
| 600 ms | - 261 ms | - 247 ms | inconsistent |
| 800 ms | - 361 ms | - 324 ms | inconsistent |
| 1000 ms | - 461 ms | inconsistent | inconsistent |

**Table 1: Group delays for varying feedback delays.** Expected group delays, or lag/lead times, from Eq. (4), average simulated group delays, and average experimental group delays, in dependence of the given feedback delay $\tau$. Positive group delays reflect that the stylus trails the target, negative that the stylus anticipates the target. The coincidence of the three group delays in the first row arises from a fit of the only free constant in model (6), the feedback gain $c_0 = 1/\tau_0$, to data with a given feedback delay of 20 ms. Simulated average values are based on 100 repeats of the trajectory for $x$ with varying initial conditions, and experimental averages are based on 10 participants [11]. The small deviations between expected and simulated group delays arise from the fact that the former are predicted from Eq. (4), which holds for small frequencies only, whereas the latter are obtained from numerical simulations of Eq. (1), which contain all frequency components of the data.

Specifically, three delay-dependent dynamic regimes are to be expected from model (6):

I) $0 < \tau \leq \tau_0$ : The group delay is non-negative. This means, the stylus position will either synchronize with or trail the target position after some transient time.

II) $\tau_0 < \tau < \tau_{crit}$ : Stable anticipation after some transient time.

III) $\tau \geq \tau_{crit}$ : Instability or inconsistent tracking.

The value of $\tau_{crit}$ depends on the properties of the frequency response function. Specifically, the larger the delay, the less the assumption for the validity of Eq. (4), namely the small frequency approximation, becomes justified, and dispersion or even oscillatory instabilities occur. This will become more apparent in the analytic group delay graphs shown in the next section.

**3 Results**

This section first describes further numerical simulations of model (1) and then a comparison with experimental data of a delayed manual tracking experiment. An interpretation of the model with respect to the observations (i) to (v) will be made in the Discussion.

In Fig. 1a sections of the input $x$ and the output $y$ are shown for one set of numerical simulations of Eq. (1) with the parameters of Eq. (6), i.e., $a = c = c_0 = 1/\tau_0 = 12.82$ s$^{-1}$, $b = 2c_0$. It is evident that *for a feedback delay of 400 ms (middle column), the response y (red) anticipates the chaotic input x (black) on average.* Like in [11], this is verified with the CCF between $x$ and $y$, which has a maximum of 0.99 at -151 ms. Further, the group delay $\delta_0$ via Eq. (4) is -161 ms (Fig. 1b, red circles). Fig. 1b also shows the estimated and analytic phase of the frequency response function. For small and intermediate feedback delays (20 and 400 ms) it has a frequency band with NGD throughout the frequency range that contains most of the power of the data $x$ (Fig. 1c). However, for very large feedback delay of $\tau = 1000$ ms, the response signal power spectrum shows some amplified components (Fig. 1c, third column). These are caused by a pole of the frequency response function that was present for the two other cases, too, but now appears in the frequency domain of the input signal $x$ (Fig. 1b, third column).

Fig. 2 shows experimental data and power spectra for a single representative subject of the experiment [11]. The observed delays as measured by the maxima of the CCFs are for $\tau = 20$ ms: CCF(+20 ms) = 0.99; for $\tau = 400$ ms: CCF(-200 ms) = 0.94. In the latter case, the subject anticipates the target with the maximum anticipation time that can be achieved with delay-induced NGD, which is given by Eq. (5) asymptotically by $\tau/2 = 200$ ms. The response signal power spectra show a similar behavior as for simulated data, namely that for $\tau = 1000$ ms some frequency components are amplified. This is caused by a pole of the frequency response function, which moves into low-frequency areas for large feedback delays (compare with Fig. 1b, third column).

All simulations in the table and the figures were performed with MATLAB R2015a (The MathWorks, Inc., Natick, MA). Equation (1) was numerically simulated with a Runge-Kutta scheme of 4$^{th}$ order, with an input described by a chaotic oscillator, and of the 160 s long simulations, the first 80 s were discarded [11].

**4 Discussion**

4.1 Modeling the observations of the manual tracking experiment

In the following, each of the observations (i) to (v) listed above will be interpreted by the one-parameter delay-induced NGD model (6).

(i) The actually measured anticipation time usually is significantly smaller than the feedback delay: This observation can be explained by our analytic result that the maximum achievable anticipation time cannot exceed half of the feedback delay time, Eq. (5).

(ii) Participants predict by an amount relative to the given feedback delay, not by the greatest of their ability: This observation can be explained by Eq. (4), which shows that the group delay depends on the feedback delay, if the other parameters are kept constant. The subject's potential ability has no influence.

(iii) The measured anticipation time roughly increases linearly with the given feedback delay for up to a certain critical delay where tracking performance becomes low: For fixed parameters $a$ and $c$, the linearity again is given by Eq.

(4). The deterioration of performance at a feedback delay $\tau_{crit}$ was shown to be related to the properties (poles) of the frequency response function for increasing feedback delays.

(iv) Tracking performance as measured by CCF decreases with feedback delay up to that critical delay: Tracking performance is directly related to the constancy of the NGD over the relevant signal components. If the group delay becomes skewed, as in Fig. 1, right column, different frequency components of $x$ are transferred differently, and the output signal $y$ is distorted relative to the input. In turn, this affects the CCF.

(v) For small feedback delay, the target is trailed by the stylus: With fixed zero lag time $\tau_0$ in model (6), this somewhat paradoxical behavior occurs for feedback gains fulfilling the condition $c_0\tau < 1$ or, equivalently, $\tau < \tau_0$. However, for sufficiently small feedback delay $\tau$ this condition can always be fulfilled. In fitting the feedback gain to trailing data, we implicitly enforced this condition (via $c_0 \times 20$ ms = 0.26 < 1) and have found a value that both allows for reactive, or trailing, and predictive, or anticipatory, behavior, depending on the feedback delay. This particular model prediction that for sufficiently small feedback delays predictive behavior is replaced by reactive behavior gives us most confidence for the validity of the model. Also, this prediction provides a clear signature of the involvement of delay-induced NGD in dynamical data in general, which can be tested in experiments.

In summary, our extremely parsimonious model (6) explains the experimental observations satisfactorily. In particular, we have found a lower bound of the feedback delay for anticipatory tracking behavior (given by condition $\tau = \tau_0$) and an upper limit determined by the properties of the frequency response function. The latter is not a strict limit but rather affects tracking performance gradually by signal distortion, when the frequency dependent NGD is not constant anymore in the main frequency range of the data, causing mixing of different frequency components, also known as dispersion. Whereas in the experiment, $\tau_{crit}$ has a value somewhere between 400 and 600 ms, in our simulations it is between 900 and 1000 ms. This discrepancy is not explained by our model. An explanation would have to include modeling *subject-dependent* upper performance limits [2], which lies beyond the present capabilities of this mechanistic model.

4.2 Causality and anticipatory synchronization

Although counterintuitive, it stands to reason that the phenomenon of NGD does not violate causality [12,19]. Its performance depends on the choice of parameters and the data, and for improperly chosen parameters and data that do not fulfill the specified stationarity and smoothness criteria it might not be predictive or cause oscillatory instabilities [15-17].

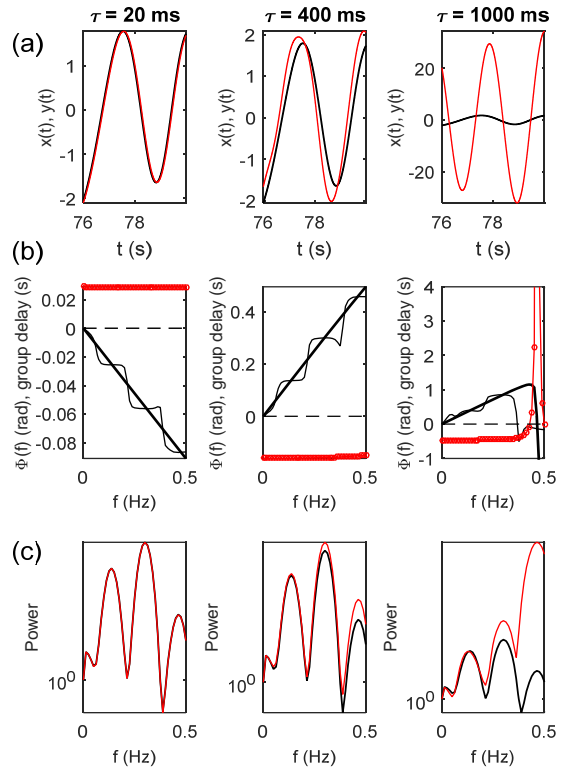

**Fig. 1: Simulation - Time series, frequency response phases, and power spectra.** The three columns each represent one of the three dynamic regimes: The first column is for a feedback delay of $\tau = 20$ ms (reactive or trailing behavior), the second column for $\tau = 400$ ms (predictive or anticipatory behavior), and the third column for $\tau = 1000$ ms (inconsistent behavior).

(a) Simulated drive $x$ (black thick graph) and response $y$ variables (red graph). Shown are the last 4 s of the 80 s that were modeled. The corresponding average lead/lag times are provided in Table 1, third column.

(b) Analytic phase, Eq. (2), (bold) and estimated phase from the simulation (thin line), as well as the analytic group delay Eq. (3) (red circles connected with lines). In the trailing case, the group delay is positive (expected value = +29 ms) and approximately constant throughout the shown frequency range; in the anticipatory case it is negative (expected value = -161 ms) and a deviation from constancy already is visible, and in the inconsistent case it is highly variable (expected value = -461 ms).

(c) Power spectra, restricted to frequencies that contain most of the power of the chaotic drive signal $x$ (black thick graph), and for the response signal $y$ (red graph). In the trailing case, power spectra are approximately identical, indicating only weak signal distortion. (The trailing effect itself does not affect the power spectrum.) In the anticipatory case some higher frequency components of $y$ are amplified, indicating a slight signal distortion; in the inconsistent case, the response is significantly distorted and tracking performance is deteriorating.

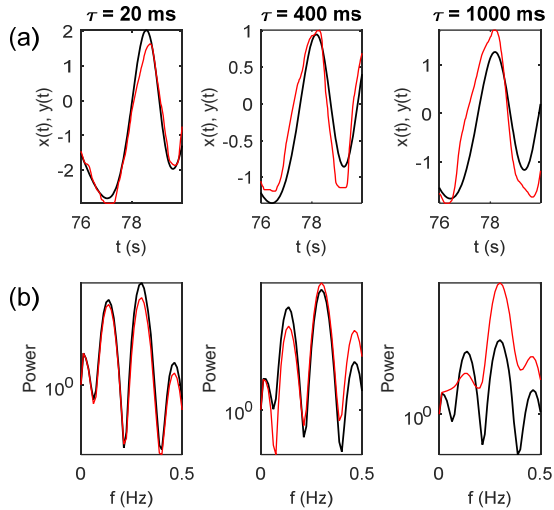

**Fig. 2: Experiment - Time series and power spectra.** The same as Fig. 1a, c, but for the experimental, normalized, data of a single subject (Subject 3, [11]). The corresponding average lead/lag times are provided in Table 1, fourth column. Frequency response functions are not shown here because model parameters are not defined for subject data

These experiments have been interpreted in a qualitative way before [11] in terms of anticipatory synchronization [19-21]. Some of the observations described here could be explained with that concept, too. Whereas anticipatory synchronization and delay-induced NGD models are related [14], the latter concept does not require a model of the driving system and is thus more parsimonious. Further, the striking difference between given delayed feedback time and actual anticipation time cannot be explained by anticipatory synchronization but follows naturally from the delay-induced NGD concept, as well as the observation of trailing behavior for small feedback times.

4.3 General model implications

Model (1), and in particular the one-parameter version of it, Eq. (6), describe the change of the stylus coordinate *y* just as a linear combination of the two other variables *x* and $y_\tau$, or as a linear filter. The cognitive task of delayed target tracking then reduces to a realization of this simple filter. In particular, the model does not require an intrinsic time delay, or memory.

Model (6) has only one parameter, $\tau_0$, which we have called the zero lag time because it determines the feedback delay for which subjects would exactly synchronize with the target without a lag or lead. The zero lag time has been kept constant for all feedback delays. It means that the model does not require an adaptation of $\tau_0$ to the feedback delay. The zero lag time has been estimated as 78 ms for a particular subject but is probably subject dependent and could contribute to the inter-subject variability observed in the tracking experiment. These model predictions could be tested in experiments with continuously changing feedback delays.

The prediction horizon Eq. (5) is universal as it is given by the mathematical properties of the delay-induced NGD frequency response function (2). Therefore, it can be used as a signature of delay-induced NGD in experiments. We note in passing that the cascading of an anticipatory system, i.e., feeding its output into another NGD system, might enable larger total prediction horizons. This would require, however, more complex circuitry and is in general less stable [15,19,22-24].

It would be very interesting if also motor experiments *without* delayed feedback can be described by this model; in this case, our model would become a *neuronal* model in which the feedback delay would be caused by internal synaptic or transmission delays [14,25]. This would lead to a description of motor control by NGD as an alternative to task-specific internal models. It might be worthwhile to augment existing neuronal models based on anticipatory synchronization [26-30] with delay-induced NGD in order to account for these observations.

The proposed mechanism is actually model free in the sense that it does not depend on the model of the trajectories to be tracked. For example, for the chaotic data of [11] it would not be required that subjects learn the chaotic dynamics beforehand. In the end, it is just the smoothness of the signals that enables prediction, independent of the underlying specific dynamics, which even can be of smoothed stochastic origin. This has been termed "anticipatory relaxation dynamics" [14]. Therefore, in order to corroborate our hypothesis that NGD and in particular delay-induced NGD plays a role in human motor control, we propose a delayed manual tracking experiment that uses smooth random [14] or otherwise unpredictable [7] signals whose dynamics cannot be learned by subjects. It has been shown that the delay-induced NGD concept is able to predict those signals, too. Such an experiment cannot be described by the entirely deterministic concept of anticipatory synchronization and could demonstrate that humans can predict random, but smooth, signals, as they might be perceived from natural sources, by utilizing NGD.

4.4 Conclusion

The delay-induced NGD model (1) accurately reproduces both trailing and anticipatory target tracking observed in a feedback-delayed manual tracking experiment. These reactive and predictive dynamics appear as two facets of the same physical mechanism, group delay. Group delay is positive for small feedback delay and negative for larger feedback delay, causing reactive and predictive dynamics, respectively.

Although the NGD utilized here is delay-induced, in contrast to most other prediction schemes our model does not require a memory of past signal values, only of past predicted states. This makes it an interesting candidate for modeling other neuronal prediction mechanisms, as the neuronal network does not have to provide memories of the input signals themselves but only of already internalized states.

[1] K. U. Smith, *Delayed Sensory Feedback and Behavior* (W.B. Saunders Co., Philadelphia, 1962).
[2] H. Gerisch, G. Staude, W. Wolf, and G. Bauch, Hum Factors **55**, 985 (2013).
[3] A. Alvarez-Aguirre, N. van de Wouw, T. Oguchi, and H. Nijmeijer, IEEE T Contr Syst T **22**, 2087 (2014).
[4] P. Tass, A. Wunderlin, and M. Schanz, J Biol Phys **21**, 83 (1995).
[5] P. Tass, J. Kurths, M. G. Rosenblum, G. Guasti, and H. Hefter, Phys Rev E **54**, R2224 (1996).
[6] W. Just, H. Benner, and E. Schöll, Adv Solid State Phys **43**, 589 (2003).
[7] R. C. Miall and J. K. Jackson, Experimental Brain Research **172**, 77 (2006).
[8] J. G. Milton, J Neural Eng **8**, 065005 (2011).
[9] J. G. Milton, A. Longtin, A. Beuter, M. C. Mackey, and L. Glass, J Theor Biol **138**, 129 (1989).
[10] A. Beuter, J. G. Milton, C. Labrie, L. Glass, and S. Gauthier, Exp Neurol **110**, 228 (1990).
[11] N. Stepp, Exp Brain Res **198**, 521 (2009).
[12] M. W. Mitchell and R. Y. Chiao, Phys Lett A **230**, 133 (1997).
[13] T. Nakanishi, K. Sugiyama, and M. Kitano, Am J Phys **70**, 1117 (2002).
[14] H. U. Voss, Phys Rev E **93**, 030201(R) (2016).
[15] M. Kandic and G. E. Bridges, Prog Electromagn Res **134**, 227 (2013).
[16] S. Bernard, J. Belair, and M. C. Mackey, Discrete Cont Dyn-B **1**, 233 (2001).
[17] O. Calvo, D. R. Chialvo, V. M. Eguiluz, C. Mirasso, and R. Toral, Chaos **14**, 7 (2004).
[18] S. Ruan, in *Delay Differential Equations and Applications*, edited by O. Arino (Springer, Berlin, 2006), pp. 477.
[19] H. U. Voss, Phys Lett A **279**, 207 (2001).
[20] H. U. Voss, in *2nd caesarium - Coupling of Biological and Electronic Systems*, edited by K.-H. Hoffmann (Springer, Bonn (Germany), 2002), pp. 119.
[21] H. U. Voss, Phys Rev E **61**, 5115 (2000).
[22] H. U. Voss, Phys Rev Lett **87** (2001).
[23] C. Mendoza, S. Boccaletti, and A. Politi, Phys Rev E **69**, 047202 (2004).
[24] A. Baraik, H. Singh, and P. Parmananda, Phys Lett A **378**, 1356 (2014).
[25] H. U. Voss, arXiv:1601.07534 (2016).
[26] M. Ciszak, F. Marino, R. Toral, and S. Balle, Phys Rev Lett **93**, 114102 (2004).
[27] F. S. Matias, P. V. Carelli, C. R. Mirasso, and M. Copelli, Phys Rev E **84**, 021922 (2011).
[28] F. S. Matias, P. V. Carelli, C. R. Mirasso, and M. Copelli, PloS one **10**, e0140504 (2015).
[29] T. Pyragiene and K. Pyragas, Nonlinear Dynam **74**, 297 (2013).
[30] T. Pyragiene and K. Pyragas, Phys Lett A **379**, 3084 (2015).